\documentclass[twocolumn]{revtex4}
\usepackage{epsfig}
\usepackage{times}

\begin{document}

\title{Accuracy of minimal and optimal qubit tomography for finite-length
  experiments} 

\author{Alex Ling}
\author{Ant{\'\i}a Lamas-Linares}
\author{Christian Kurtsiefer}

\affiliation{Centre for Quantum Technologies and Department of Physics,
  National University of Singapore, Singapore 117543}

\date{\today}

\begin{abstract}
Practical quantum state tomography is usually performed
by carrying out repeated measurements on many copies of a given state. The
accuracy of the reconstruction depends strongly on the dimensionality of the
system and the number of copies used for the measurements. We investigate the
accuracy of an experimental implementation of a minimal and optimal tomography
scheme for one- and two-qubit states encoded in the polarization of photons. A
suitable statistical model for the attainable accuracy is introduced.
\end{abstract}


\maketitle

\section{Introduction}        
The accurate characterization of quantum states and their evolution is central
to quantum information processing. Quantum tomography or state estimation
attempts to extract as much information about the state as possible out of
physically realizable measurements on an ensemble of identically prepared
copies. Qubits are the simplest carriers of quantum information and thus the
estimation of ensembles of multi-qubit states have received particular
attention.

A full quantum state estimation for a system composed of $n$ qubits demands
the determination of $2^{2n}-1$ real-valued parameters.
The accuracy in estimating
each of these parameters decreases rapidly with the system size $n$ to an
extent that it becomes difficult to assess the operation of the system even
for a moderate number of qubits \cite{haeffner:05}. In practice, the
estimation ensemble is limited, hence it is critical to maximize
the amount of information extracted out of each copy.

Substantial work has been targeted towards understanding optimal methods for estimating quantum states under different constraints \cite{optimal,collective,separable,latorre,gillandmassar,ramon1,mqt,sims07}, 
partly also motivated
by understanding the information leakage to an eavesdropper in a quantum
communication scenario \cite{sprot}.
The notion of `optimal', however, is ambiguous and depends on the constraints
imposed. It might refer to the complexity of the measurement scheme (number of
outcomes, projection measurements vs POVMs), to the number of copies consumed
for a particular confidence level, or to both. Most theoretical treatments
compare the most general collective measurements to projective measurements on
individual qubits, and are only valid in the asymptotic limit of infinite
copies. This leaves the experimentally more relevant case of small number of
copies where POVMs can be performed on each qubit as a largely unexplored
regime.

Experimentally, only finite POVMs are feasible, and it is often highly desirable to have a minimal number of outcomes (smallest possible number sufficient to characterize the target Hilbert space). Additionally, it is often not possible to change the measurement from copy to copy, so exactly the same POVM is performed on all members of the ensemble. Thus, we want a minimal POVM that is optimal given its fixed number of measurement outcomes.  Such a minimal and optimal POVM was described by \u{R}eh\'{a}\u{c}ek {\it et al.} for estimating qubit states \cite{mqt}. 

In this paper, we investigate how the accuracy of direct state estimation for the minimal and optimal POVM described in \cite{mqt} changes as we increase the number $N$ of copies for one and two-qubit states.  Such a scenario is regularly encountered in applications where copies are precious resources, yet the state must be estimated with some accuracy.  In this case, it is desirable to know the number of copies that must be sacrificed to reach a certain level of accuracy.

We implement qubit states using the polarization degree of freedom in single photons, and the POVM using polarimeters as in \cite{optpolarimeter,opttomography}, and briefly review the experimental system in  Section~\ref{sec:minitom}. In Section~\ref{sec:statmodel}, we describe a statistical model that gives the accuracy of state estimation for a test state, given any number $N$ of copies.  We experimentally verify the statistical prediction by observing maximally polarized one-photon states in Section~\ref{sec:experiment}.  In Section \ref{sec:twoqubit}, we present observations for similar behavior with two-photon states, as two or more  such polarimeters can be used to reconstruct entangled multi-qubit states \cite{opttomography}.  Our results suggest a scaling law governing the accurate reconstruction of all multi-qubit states with this method. We conclude in Section~\ref{sec:conclusion}.

\section{Minimal and optimal polarimetry by photon counting}
\label{sec:minitom}

\begin{figure}
\centerline{\epsfig{file=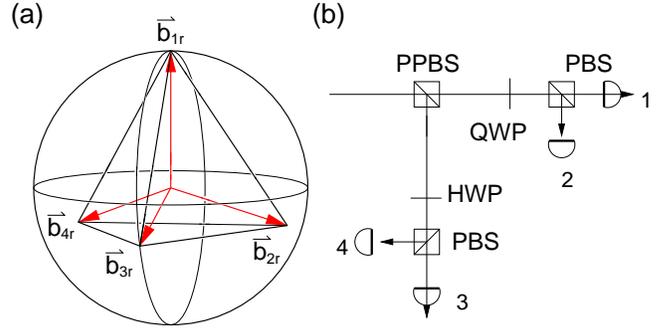, width=86mm}} 
\caption {\label{optimal}Optimal polarimetry. (a) To extract the Stokes
  parameters from an ensemble of photons, polarization measurements are
  carried out corresponding to four measurement states $\vec{b}_j$, which form
  a tetrahedron on the Pointcar\'{e} sphere.
  (b) Experimentally, four photon counting detectors are
  used, each associated with the outcome of one of the measurements.
  The light is divided between the four detectors by a partially polarizing
  beam splitter (PPBS), half (HWP)- and quarter (QWP) wave plates, and
  polarizing beam splitters (PBS) \cite{optpolarimeter}.} 
\end{figure}

There is a close connection between the polarization state of light in
classical optics, and the state of a qubit implemented via the polarization
states of photons \cite{james:01}. The macroscopic polarization
of light can be thought of as formed by an ensemble of equally prepared
photons, where a photon represents the minimum amount of detectable light. A
Stokes vector, $\vec{S}=(1,S_1,S_2,S_3)$, describes the macroscopic
polarization of light \cite{stokes}, and also is a representation of the
density matrix of the qubit state. In classical optics the process of
determining the polarization state is called polarimetry \cite{dafdp}, and its
techniques are often similar to those used in quantum tomography
\cite{optpolarimeter}.

The main principle of minimal polarimetry is to use the minimum
number of measurements to characterize the three free parameters of the Stokes
vector, $S_1,S_2$ and $S_3$.  Thus, only four measurements are required  for
complete polarimetry (the fourth one necessary only to obtain a normalization
to the total intensity or number of photons present), each corresponding to a
particular Stokes vector.
If these four Stokes vectors form a tetrahedron in the 
Poincar\'{e} sphere (Figure \ref{optimal}a), then the set of four measurements
is minimal and optimal \cite{latorre}, where optimal means that the
measurements extract the maximum amount of information from each photon that
is detected. 

The optimal polarimeter works by partitioning the test ensemble between four
detectors (see Figure \ref{optimal}b).
A four-by-four instrument matrix $B$, where each row is one of the measurement
vectors, completely characterizes the polarimeter \cite{dafdp}.  
Detector event distribution and input polarization are connected by the linear
form 
\begin{equation}
\label{eq:lintom}
\vec{I}=B\cdot \vec{S}\,,
\end{equation}
where $\vec{I}=(I_1,I_2,I_3,I_4)$, with the relative number $I_n$ of
events registered by detector $n$.  Knowing $B$, the observed
distribution among the detectors is used to perform linear reconstruction of
the input Stokes vector.

We implemented the polarimeter with avalanche photo-diodes which are sensitive
to single photons.  In this way, we can keep track of the photon distribution
between the detectors as each copy of the qubit state is detected.

\section{A statistical model for predicting the accuracy of state estimation}
\label{sec:statmodel}

A qubit state can only be determined with an infinite number of copies.
Experimentally, a good estimate of the state is achieved by collecting an
asymptotically large number of copies.  We refer to this estimate as the
asymptote state.  This estimate presents the best possible guess we can make
about the unknown input state.

\begin{figure}
  \centerline{\epsfig{file=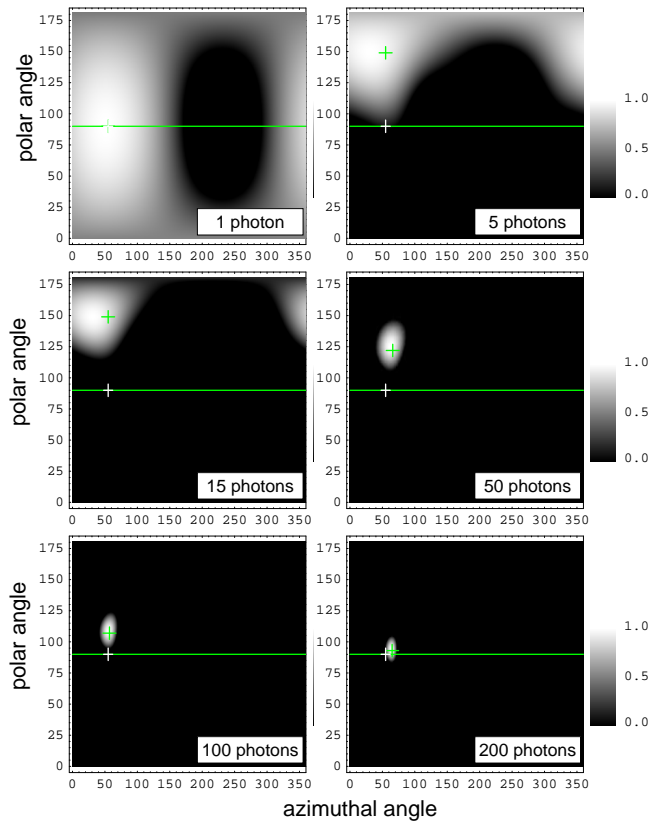, width=86mm}}
  \caption{\label{converge}
    Projections of the reconstructed probability distribution on the
    Poincar\'{e} sphere surface for a series of measurement outcomes form a
    polarization state ensemble. The horizontal line in the center of each panel represents
    linearly polarized states.  Light crosses mark the prepared polarization
    state, and a darker cross marks the estimated (assumedly pure) state.} 
\end{figure}

The finite number of copies $N$ limits the accuracy in the estimated state.
Although we use the event distribution to directly
obtain an estimated state through (\ref{eq:lintom}), 
every distribution has a deviation from the asymptotic value, and cannot 
be attributed with absolute certainty to a single Stokes vector $\vec{S}$.
At best, we can assert that the estimated state is compatible with the
asymptote state, given the estimated counting uncertainty due to
Poisson distribution (assuming independent sequential detection events).

To determine the accuracy of state reconstruction, we consider only specific
input states. Since we find experimentally that the accuracy of the worst and
best 
reconstruction cases are quite close, with the counting errors making the two
cases indistinguishable, we avoid an analysis of averaging over all possible
input states. 

The accuracy of an estimated state obtained with a finite ensemble must
eventually converge to the accuracy of an asymptotic estimate.  Being able to
determine the accuracy is useful as it provides a confidence level to an
estimated state, based on the number $N$ of detected copies.
As a quantitative measure of estimation accuracy, we use the average
of the trace distance $D=\frac{1}{2}{\rm tr}|\rho_a - \rho_e|$ of an estimated
state $\rho_e$ from the asymptote state $\rho_a$ \cite{distance}, where the
averaging occurs over a few consecutive experiments with a fixed number of
detection events each.

To model the expected average trace distance for our experimental setting, we
must find all the possible ways to distribute $N$ detection events between
four detectors.  For each partition pattern 
$k=(n_1,n_2,n_3,n_4)$, there are a total of $c_k=N!/(n_1! n_2! n_3! n_4!)$
compatible detector sequences. Each sequence in a pattern $k$ results in the
same reconstructed state $\vec{S_k}$ and consequently same trace
distance $D_k$ from the asymptote state.  The probability of each sequence is
given by
\begin{equation} 
p_k=p_{1}^{n_1} 
\cdot p_{2}^{n_2} \cdot p_{3}^{n_3} \cdot p_{4}^{n_4}\,,
\end{equation}
where $p_{j}$ is
the probability that an input photon will arrive at detector $j$. 
Thus, the weighted average $\tilde{D}$ of the trace distance
\begin{equation}\label{eq:statmodel}
\tilde{D} =\sum_k c_k\cdot p_k\cdot D_k
\end{equation}
represents the accuracy of the estimated state given $N$ copies. 

\section{Reconstruction of one-qubit states}
\label{sec:experiment}
\subsection{Maximally polarized or pure one-qubit states}
In a first experiment, we consider linear state reconstruction
for an increasing number of detected events, converging to an
asymptote state.  At each stage, we also perform a likelihood estimation to find
a region of states that are compatible with the observed photon
distribution.  The size of this region of states can be interpreted
qualitatively as the uncertainty in our estimate.

We used spontaneous parametric down conversion (SPDC) in a
non-collinear setup similar to \cite{kwiat95} to generate 
heralded single photons \cite{singlephoton} with a controllable polarization
state. This 
allows us to select a well defined ensemble of carriers of the qubit
state, virtually unaffected by accidental counts and background noise.
Photons were then prepared in a maximally polarized state using a polarization
filter and a set of wave plates \cite{arbitrarystates}.
We chose to test the polarization state corresponding to the measurement
vector $\vec{b_{1r}}=(1, \sqrt{\frac{1}{3}},\sqrt{\frac{2}{3}},0)$ of the
polarimeter, because states aligned with the tetrahedral measurement
directions are estimated with the 
poorest accuracy.  Vectors anti-aligned with the tetrahedron directions have
the best estimation accuracy due to their restricted photon distribution pattern
\cite{mqt}.  The worst case scenario should give us a lower bound of
the POVM performance; other states should have at least the same accuracy.  

We detected 200 copies prepared in this state.  For each additional copy, we
obtained a state  by linear reconstruction under the constraint of the nearest
physical 
state.  Concurrently, the likelihood region is determined
for a given number of detection events. A projection of both the estimated
state and the likelihood region on the surface of the Poincar\'{e} sphere is
shown in Figure \ref{converge} for a selected number of cumulative detection
events. Initially, for a small number of available detector outcomes, the
estimated state fluctuates strongly, and the likelihood region is large.
As the accumulated number of copies increases, the estimated state
begins to approach the asymptote state, and the
likelihood region shrinks, indicating a reduction of uncertainty in
the estimated state.

\subsection{Accuracy as a function of the detected number of copies}
The results of the previous section revealed qualitatively the convergence of
an estimated state to the asymptote state.  To study the convergence
quantitatively, we determine the average trace distance for a
given number of detected copies. 

We selected three test states: the tetrahedron state $\vec{b_{1r}}$,
its conjugate $-\vec{b_{1r}}$, and the completely unpolarized state
$\vec{S}=(1,0,0,0)$.  The latter is obtained by collecting light from one arm
of the SPDC source without any polarization filters, and is a test for our model
for mixed states.  The two maximally polarized states represent
the worst and best cases in estimating pure states. 

For each test state,  a very large number of heralded copies (several $10^5$)
was detected to obtain an approximation to the asymptote state. 
Next, for each incrementally obtained
copy, an estimated state and the corresponding trace distance to the asymptote
state was determined.  No restrictions were used
in obtaining these estimates.  Such finite sized collections were repeated 40
times, from which we obtained the average trace distance for each additional
copy that was measured. 

\begin{figure}
  \centerline{\epsfig{file=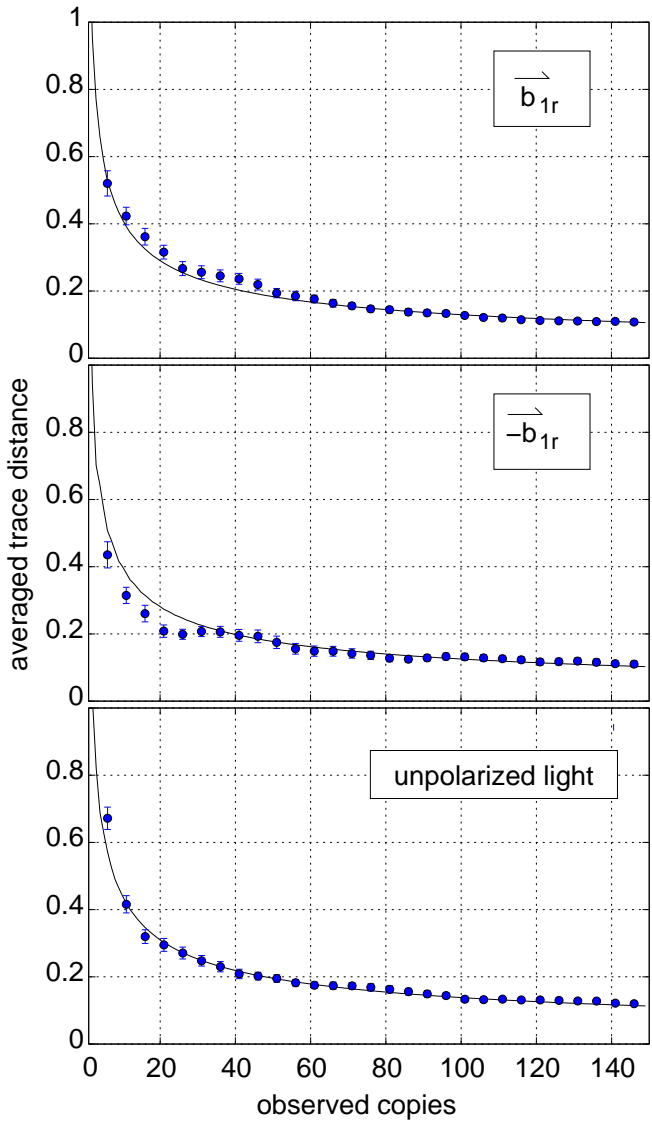, width=86mm}}
  \caption{\label{fig:combo}
    Trace distance between estimated and asymptotic states for
    three different input states. The experimental values 
    represent an average over 40 runs with 150 detection events each.
    Solid curves represent the statistical model (\ref{eq:statmodel}).
  } 
\end{figure}

Selected subsets for the different test states are shown in Figure
\ref{fig:combo}.  The average trace distance predicted via
(\ref{eq:statmodel}) is shown 
by the solid line.  The accuracy of the experimental POVM is consistent with
the statistical model for both polarized and unpolarized light, and most of
the increase in accuracy occurs within the first 100 detected copies.
Such a graph can be useful for predicting the average accuracy of any
estimation from a finite ensemble of copies.  Both analytical and experimental
results are consistent with simulations that have been previously reported \cite{mqt,sims07}.
We note that our statistical model is also compatible
with the prediction in \cite{mqt} concerning different accuracy levels for
different states.  However, from any single experimental run, the counting
uncertainty make the accuracy levels compatible. 
Thus, the accuracy of state estimation by the optimal POVM is in practice the
very similar for all states.  

\begin{figure}
  \centerline{\epsfig{file=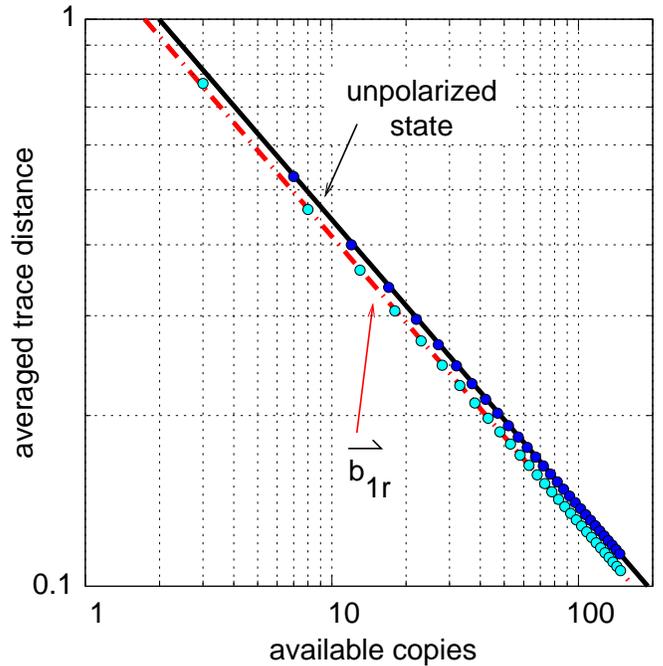, width=86mm}}
  \caption{\label{fig:loglogcombo}
    Average trace distance for two input states using only the results of the
    statistical model (points). Lines are fits to the power law model
    (\ref{eqnD}).  } 
\end{figure}

The results from the statistical model may be analyzed further to reveal how
the average trace distance $\tilde{D}$ diminishes with the number of available
copies $N$. A logarithmic representation of the analytical results (see
Fig.~\ref{fig:loglogcombo}) suggests a power law of the form
\begin{equation}\label{eqnD}
  \tilde{D}=\frac{a}{N^c}\,.
\end{equation}

The values for parameters $a$ and $c$ extracted from a least-squares fit to
the analytical results for some test states are presented in
Table~\ref{table:fitparams}.
The exponent $c$ is close to 1/2, as perhaps expected form a
simple counting statistics argument.
Parameter $a$ in (\ref{eqnD}) seems to represent the difficulty in estimating
a particular state; pure polarization states all lead to slightly lower values
than the completely unpolarized state.

\begin{table}[hb]
  \begin{center}
    \begin{tabular}{r|c|c|c|c}
       & unpolarized & horizontal& $\vec{b_{1r}}$ & $\vec{-b_{1r}}$ \\
      \hline
      $a$ & 1.417 & 1.312 & 1.323 & 1.288 \\ \hline
      $c$ &  0.506 & 0.505 & 0.505 & 0.506 \\
    \end{tabular}
  \end{center}
  \caption{Power law fit parameters for different test
    states\label{table:fitparams}}
\end{table}

\section{Accuracy in reconstructing a two-qubit state}
\label{sec:twoqubit}

State estimation of multi-qubit states using only separable measurements and
classical communication is readily implemented by a generalization of the
previous scheme. For a two photon state, a pair of polarimeters are used
simultaneously to determine the generalized Stokes vector; each photon in the
joint state is analyzed by a separate device  \cite{opttomography}.  By
looking at the distribution 
of coincidence counts between the polarimeters, the two-qubit state,
adequately described by a two-photon Stokes vector \cite{optpolarimeter,2psp},
can be reconstructed.

The photon pairs were prepared in the  Bell state
$|\Psi^+\rangle = \frac{1}{\sqrt{2}}(|HV\rangle + |VH\rangle)$ with the same
SPDC source as in the one-qubit experiments. Several $10^5$ copies of the
two-qubit state were 
detected to obtain an estimate of the asymptote state.  Then, five sets, each
with 5000 detected pairs, were analyzed.  Within each set,
an estimated state for each incremental pair was obtained, together with its
trace distance to the asymptote state.  Trace distance values for every
cumulative event number were then averaged over those five sets.

To compare the resulting trace distances with the one-photon tests, we
introduce a normalized trace distances  $\tilde{D_n}=\tilde{D}/(2^{2n} - 1)$
as the trace distance of the (multi)-partite system divided by the number of
free parameters $n$.
This normalization attemps to capture the exponential increase in the number
of parameters to be determined as the dimensionality increases.

\begin{figure}[t]
  \centerline{\epsfig{file=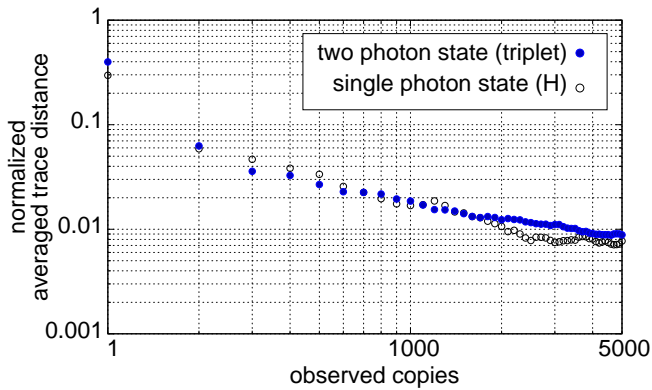, width=86mm}} 
  \caption{\label{combob}
    Averaged trace distance (over five experimental runs) for a finite number
    of detection events for the two-qubit Bell state $|\Psi^+\rangle$ and a
    horizontally polarized one-qubit state, normalized to the number of free
    parameters (3 for a one-photon state and 15 for the two-photon state).  }
\end{figure}

The normalized results are compared in Figure \ref{combob}. We
find that for both one- and two- photon systems, the average trace distance is
within 0.01\% after 5000 detection events, and appears to follow the same
dependence on the number of copies $N$.  This suggests that the separable POVM
reconstructs multi-qubit states with the same accuracy per copy and dimension
of the multi-qubit  state, 
providing possibly a simple scaling law to determine the number of copies
necessary to estimate a multi-qubit system to any given accuracy. It also
provides an experimentally relevant benchmark to evaluate different 
measurement strategies.

\section{Summary}
\label{sec:conclusion}
We have presented an experimental observation of the average accuracy of a
minimal and optimal POVM, taking into account the uncertainty due to a finite
number of available copies of a state.  The average trace distance
was identified as a measure of accuracy, and the reduction in trace distance
with the number of detected copies was studied. The results were compatible
with a simple model for any number of detected copies using multinomial
statistics.

The estimation uncertainty for different input states was compatible with
Poissonian counting statistics for realistic experimental 
situations.  With this view, the POVM method is able to estimate all
single-qubit states equally well. 

Furthermore, the normalized accuracy in estimating a two-qubit state appears
to follow the same scaling law as in the one-qubit case, suggesting a scaling
law for all multi-partite systems.

\section*{Acknowledgements}
We acknowledge financial support from ASTAR SERC under grant No.~052~101~0043,
and the National Research Foundation and the Ministry of Education, Singapore.

\end{document}